\documentclass[aps,preprintnumbers,eqsecnum,amsmath,amssymb,nofootinbib]{revtex4}
\usepackage{graphicx}
\usepackage[english]{babel}
\usepackage{url}
\usepackage{graphicx}
\usepackage{xcolor}
\usepackage{amsmath}
\usepackage{amssymb}
\usepackage{slashed}
\usepackage{multirow}
\begin{document}
\title{{\boldmath $B_{(s)}\to V M_X$} decays as probes of dark-matter scenarios \\
 of Belle II enhancement in {\boldmath $B\to KM_X$} decays}
\author{Alexander Berezhnoy$^{a}$, Wolfgang Lucha$^{b}$, and Dmitri Melikhov$^{a,c,d}$}
\affiliation{
$^a$D.~V.~Skobeltsyn Institute of Nuclear Physics, M.~V.~Lomonosov Moscow State University, 119991, Moscow, Russia\\
$^b$Marietta Blau Institute (MBI) for Particle Physics, Dominikanerbastei 16, A-1010 Vienna, Austria\\
$^c$Joint Institute for Nuclear Research, 141980 Dubna, Russia\\
$^d$Faculty of Physics, University of Vienna, Boltzmanngasse 5, A-1090 Vienna, Austria}
\date{\today}
\begin{abstract}
  Recently, we have shown that the hypothesis of the dark-matter (DM) origin of the
  Belle II excess events in $B\to K M_X$ decays allows for a successful description of
  the data. The DM parameters (masses and couplings) in two DM scenarios with either
  scalar (S-scenario) or vector (V-scenario) mediator have been extracted with rather
  small uncertainty from fits to the $B\to K M_X$ data. The same mechanism leads to a
  similar enhancement of the $B_s\to (\eta,\eta',\phi) M_X$ decay rates. We present a
  detailed analysis of $B\to K^* M_X$ and $B_s\to \phi M_X$ decays and show that these
  decays provide a clear probe of the S- and V-scenarios: Compared to the Standard Model,
  in the S-scenario an enhancement by a factor $\sim 1.6$ is expected, whereas the
  V-scenario leads to a much larger enhancement by a factor $\sim 3.5$.
  For the $B\to K^* M_X$ decays, we provide the expected number of events
  at Belle~II for the sample for which a rather large enhancement of
  $B\to K M_X$ decays compared to the Standard Model expectations has been reported.
\end{abstract}
\maketitle

\section{\label{Sect:1}Introduction}
Within the framework of the Standard Model (SM) of particle physics, the missing-energy $B^+$ decay $B^+\to K^+ M_X$
is enacted by the decays $B^+\to K^+\bar\nu\nu$ into a pair of neutrino and antineutrino.
This flavour-changing neutral-current (FCNC) induced process is, in the SM, forbidden at tree level and thus
only proceeds via loops (penguins and boxes). For the resulting branching ratios, rather accurate theoretical
predictions have been reported \cite{damir2023epjc,hpqcd2023,Allwicher:2023xba}:
\begin{eqnarray}
 \label{damir}
Br(B^+\to K^+\bar\nu\nu)=(4.44\pm 0.30) \times 10^{-6}, \qquad 
Br(B^+\to K^{*+}\bar\nu\nu)=(9.8\pm 1.4) \times 10^{-6}.
\end{eqnarray}
Experimentally, in 2017 Belle reported for these branching ratios, in agreement with the SM,
the upper limits \cite{belle2017}
\begin{eqnarray}
\label{belle2017}
Br(B\to K\bar\nu\nu)< 1.6\times 10^{-5}, \qquad
Br(B\to K^*\bar\nu\nu)< 2.7\times 10^{-5}.
\end{eqnarray}
In 2023, however, Belle II reported \cite{Belle-II:2023esi} a surprising observation significantly
exceeding the SM expectation,
\begin{eqnarray}
Br(B^+\to K^+\bar\nu\nu)&=&(2.3\pm 0.7) \times 10^{-5}=(5.4\pm 1.5)Br(B^+\to K^+\bar\nu\nu)_{\rm SM},
\end{eqnarray}
and, in addition, the measurement of the differential distribution for the decay $B\to K M_X$.
If eventually confirmed, these findings may be interpreted as arising from the contribution of
new physics beyond the SM. Needless to stress, lots of inequivalent new-physics scenarios
attempting to provide an explanation of the $B\to K\bar\nu\nu$ enhancement, relying~on very
different underlying concepts (among others on leptoquarks, non-universal $U(1)'$ models,
sterile neutrinos, Higgs portal, minimal flavor violation, R-parity-violating supersymmetry,
or light new particles) have been proposed
\cite{Fuentes-Martin:2020hvc,Athron:2023hmz,Bause:2023mfe,Felkl:2023ayn,Fridell:2023ssf,Wang:2023trd,He:2023bnk,McKeen:2023uzo, Altmannshofer:2023hkn, Altmannshofer:2024kxb,Bhattacharya:2024clv,Davoudiasl:2024cee,Kim:2024tsm,Marzocca:2024hua,
Bolton:2024egx, Loparco:2024olo,Zhang:2024hkn, fajfer2503,DiLuzio:2025whh,Altmannshofer:2025eor}.

Particularly appealing is the interpretation of the Belle II excess events by $B$ decays into
invisible dark-matter~(DM) particles. This possibility has been widely discussed
\cite{Abdughani:2023dlr,Berezhnoy:2023rxx,datta,Calibbi:2025rpx,He:2024iju,Hou:2024vyw,Gabrielli:2024wys,Ho:2024cwk,
 He:2025jfc,Kolay:2025jip,blm2025,Aliev:2025hyp,Ding:2025eqq,He:2025zfy,blm2026,universe,Liu:2025lbw,Kim:2025zaf, Mescia:2026xju,Bolton:2025lnb,Jahedi:2026dzy,Gartner:2026clx,Ho:2026kqp}.
Various specific scenarios have been considered, e.g., a dark photon or decays into
DM particles via some mediator field.
With respect to the last option, for the coupling of the mediator field to the SM particles
different patterns may be envisaged.

In the past, we focused on the missing-energy decays $B\to K^{(*)} M_X$
\cite{Berezhnoy:2023rxx,blm2025,blm2026,universe}, demonstrating \cite{Berezhnoy:2023rxx} that,
within~a scalar-mediator scenario (S-scenario), rigorous constraints on the differential rates
of $B\to K^* M_X$ excess events may be obtained, showing \cite{blm2025} that the S-scenario
provides an excellent description of the differential distributions of the excess
events reported by \cite{Belle-II:2023esi},
and extracting \cite{blm2025} the corresponding DM parameters.
In a similar manner, we analyzed a vector-mediator scenario (V-scenario) and
extracted the related DM parameters \cite{blm2026}.

Here, we continue to investigate the conjecture that the $B\to K M_X$ excess reflects decays
into pairs of invisible~DM fermions $\bar\chi\chi$ coupling to the top-quark
(whence dubbed a ``top-philic scenario'') via a scalar or vector mediator field~$R$:
\begin{eqnarray}
\Gamma(B\to K M_X)=\Gamma(B\to K \bar\nu\nu)_{\rm SM}+\Gamma(B\xrightarrow{R}K \bar\chi\chi).
\end{eqnarray}
Within top-philic scenarios, one and the same pattern of excess events is expected for all
decay modes, $B\to (P,V)M_X$ with $P=K,\pi$, $V=K^*,\rho$, and $B_s\to (P,V)M_X$
with $P=\eta,\eta',K$, $V=\phi,K^*$, compared to the SM predictions for these modes.
For the SM and the DM S- and V-scenarios, we give a parameter-free prediction
for $d\Gamma(B_s\to \phi M_X)/dq^2$ and show that the latter's experimental observation
provides a powerful discriminator between DM scenarios.

\section{\label{Sect:2}Standard Model and Dark-Matter contributions to $B_{(s)}\to (K,K^*,\phi)M_X$}
For the missing-energy decays $B_{(s)}\to (K^*,\phi) M_X$, we work out implications of the
conjecture that the Belle~II~excess events are due to the decay into dark-matter fermions,
generically called $\chi$, coupled to the top quark via some mediator field $R$.
No interaction of this mediator $R$ with other SM particles is implied. In this case,
the FCNC decay $b\xrightarrow{R}s\bar\chi\chi$ proceeds via the top--$W$-boson loop depicted in Fig.~\ref{Fig:1}:
\begin{center}
\begin{figure}[ht]
 \includegraphics[width=5cm]{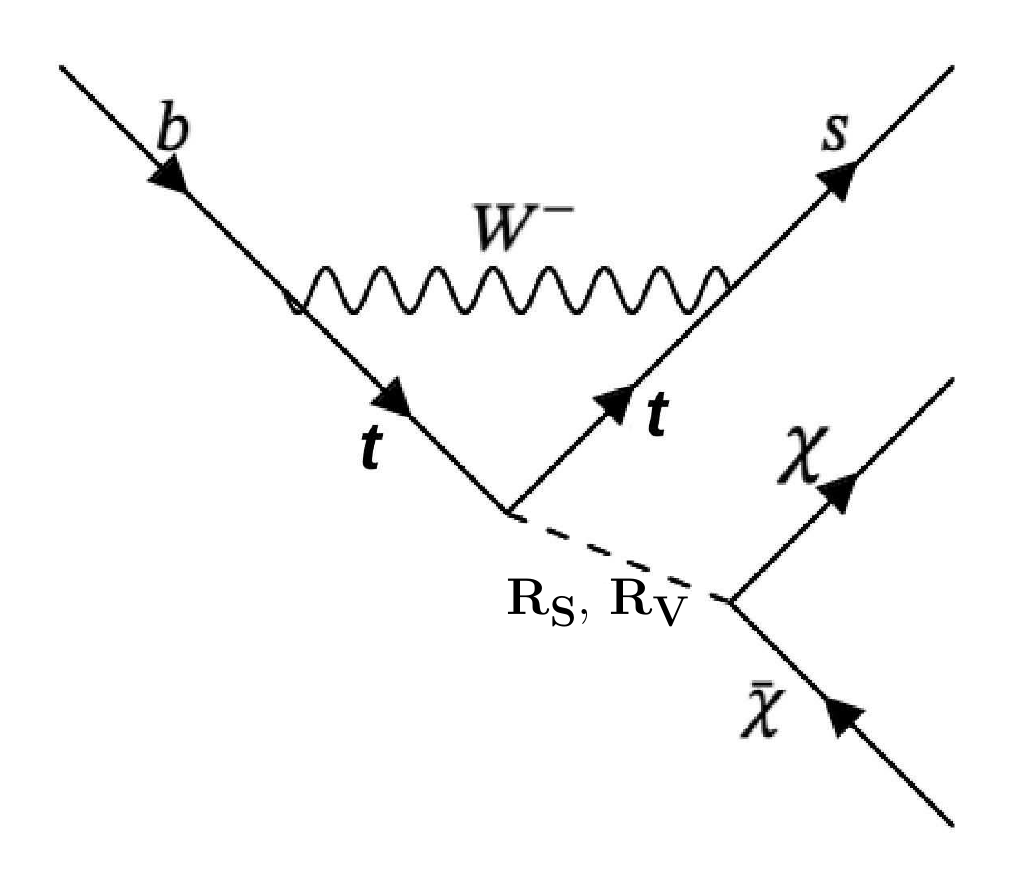}
 \caption{\label{Fig:1}Diagram describing the $b\to s\bar\chi\chi$ amplitude in a top-philic DM scenario.}
\end{figure}
\end{center}
As announced in the Introduction, we will consider both scalar (S-scenario) and vector (V-scenario) mediator fields.

\noindent
$\bullet$ 
Appendix \ref{AppendixA} provides all necessary formulas for the effective
Hamiltonians $H_{\rm eff}$ describing the $b\to s$ transition at the
scale $\mu=M_W$ (including the operator basis and the corresponding Wilson coefficients)
for the SM and the S- and V-mediator DM scenarios as well as for the resulting
expressions of the differential distributions for the missing-energy decays $B\to (P,V)M_X$
within both SM and DM scenarios.

\noindent
$\bullet$ 
Apart from the parameters introduced by the effective Hamiltonians $H_{\rm eff}$, the amplitudes
of interest involve form factors parameterizing the hadronic matrix elements governed
by the $b\to s$ transition.
Appendix \ref{AppendixB} provides a detailed discussion of these quantities and the parameterizations
of all those form factors necessary for our calculations.

With all these ingredients at hand, we can readily perform the analysis of the decay modes of interest.

\section{DM contributions to the observables}
The relevant DM parameters have been extracted by fitting the Belle II excess events within both S- and V-scenarios.
Note that, to avoid confusion, in this section we denote scalar mediators by $R_s$ and vector mediators by $R_v$,
whereas the notation $\phi$ is reserved for the vector $\phi$-meson.

\noindent
$\bullet$
For the S-scenario, the ``best'' parameter values corresponding to the global
  minimum of $\chi^2$ were reported in~\cite{blm2025}:
\begin{eqnarray}
 \label{DMS}
 M_{R_s}=2.4\pm 0.4 \mbox{ GeV},\quad \Gamma^0_{R_s}=2.9^{+1.1}_{-0.9} \mbox{ GeV},
 \quad m_\chi=0.42^{+0.2}_{-0.4}\mbox{ GeV}, \quad g_{R_s\chi\chi}\approx 6.08,\quad
 g_{bsR_s}\approx 5.6 \times 10^{-8}.\quad
\end{eqnarray}
The coupling $g_{R_s\chi\chi}$ is obtained, via Eq.~(\ref{Gamma}), from the extracted values of $M_{R_s}$
and $\Gamma_{R_s}^0$. From the total~number of $B$ mesons produced at Belle II ($N_\mathrm{tot}=3.99 \times 10^8$),
using this $g_{R_s\chi\chi}$ and the observed excess of approximately 170 events,
yields $g_{bsR_s}$.

\noindent
$\bullet$  
 For the V-scenario, using a similar algorithm, we found \cite{blm2026}
\begin{eqnarray}
 \label{DMV}
 M_{R_v}\le 3 \mbox{ GeV},\quad \Gamma^0_{R_v}=4.0^{+2.0}_{-1.5} \mbox{ GeV},
 \quad m_\chi=0.6^{+0.10}_{-0.18}\mbox{ GeV},\quad
g_{R_v\chi\chi}\approx 7.13, \quad g_{bsR_v}\approx 1.92\times 10^{-8}.
\end{eqnarray}
The theoretical decay rates $Br(B\xrightarrow{S}K\chi\chi)$ and $Br(B\xrightarrow{V}K\chi\chi)$
corresponding to these couplings turn out to~be slightly different:
\begin{eqnarray}
 & Br(B\xrightarrow{S} K M_X)=1.95\times 10^{-5},\\
 & Br(B\xrightarrow{V} K M_X)=1.52\times 10^{-5}.
\end{eqnarray}
Notice that the branching ratios are different in S- and V-scenarios. This happens for the following reason: 
The couplings $g_{bsR_s}$ and $g_{bsR_v}$ are determined by reproducing the {\it observed} number of the excess
events. The latter depends on the $q^2$-dependent experimental efficiency $\varepsilon(q^2)$
estimated in \cite{Fridell:2023ssf}.  Because of the different $q^2$-dependence of the differential
decay rates in S- and V-scenarios,
the {\it theoretical} integrated branching ratios -- i.e., without applying $\varepsilon(q^2)$ -- 
turn out to be different. 

\begin{center}
 \begin{figure}[t]
 \begin{tabular}{cc}
 \includegraphics[width=8.5cm]{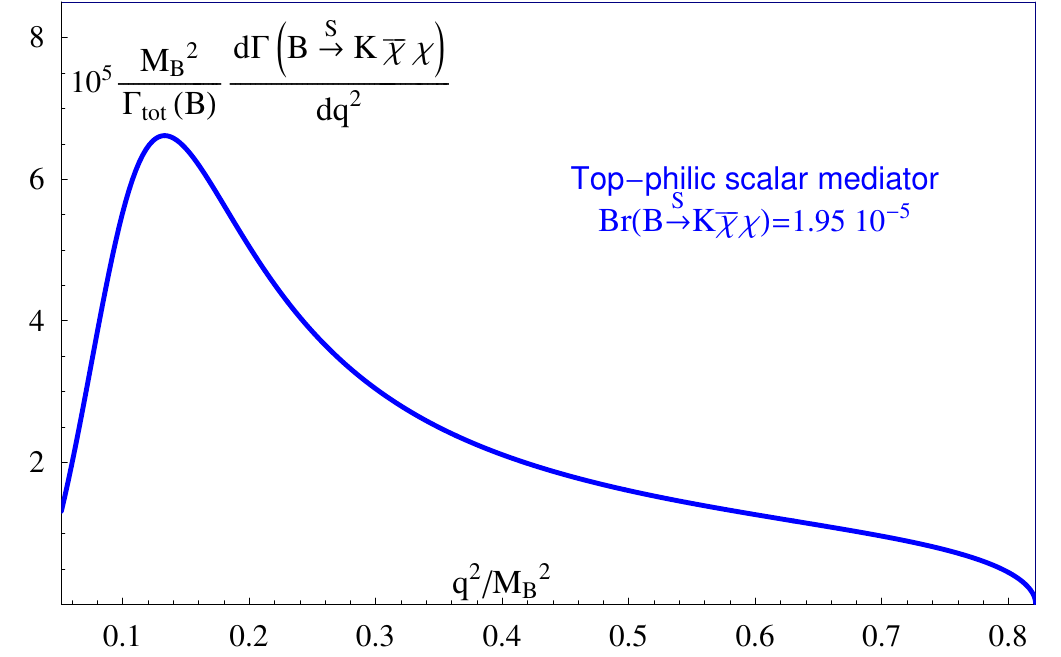} & \includegraphics[width=8.5cm]{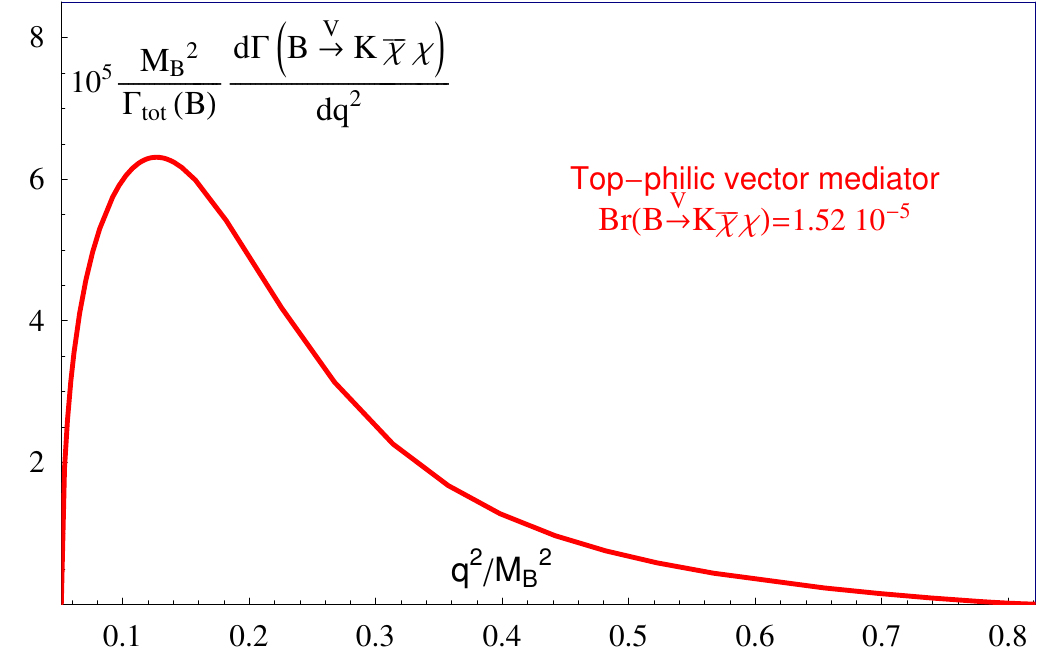}\\
 (a) & (b) \\
 \includegraphics[width=8.5cm]{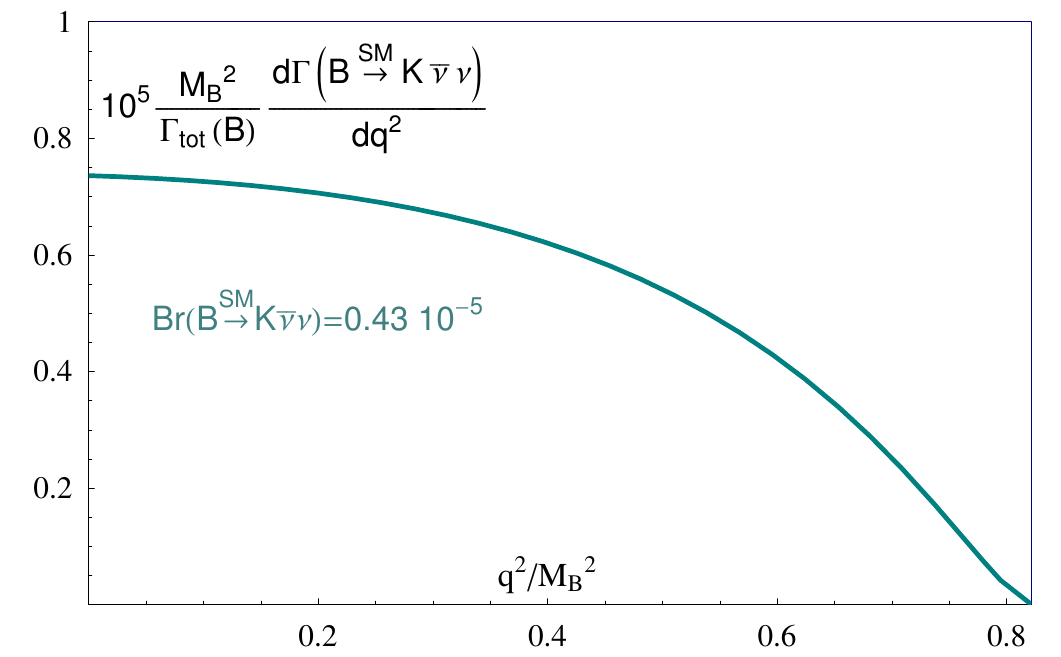} & \includegraphics[width=8.5cm]{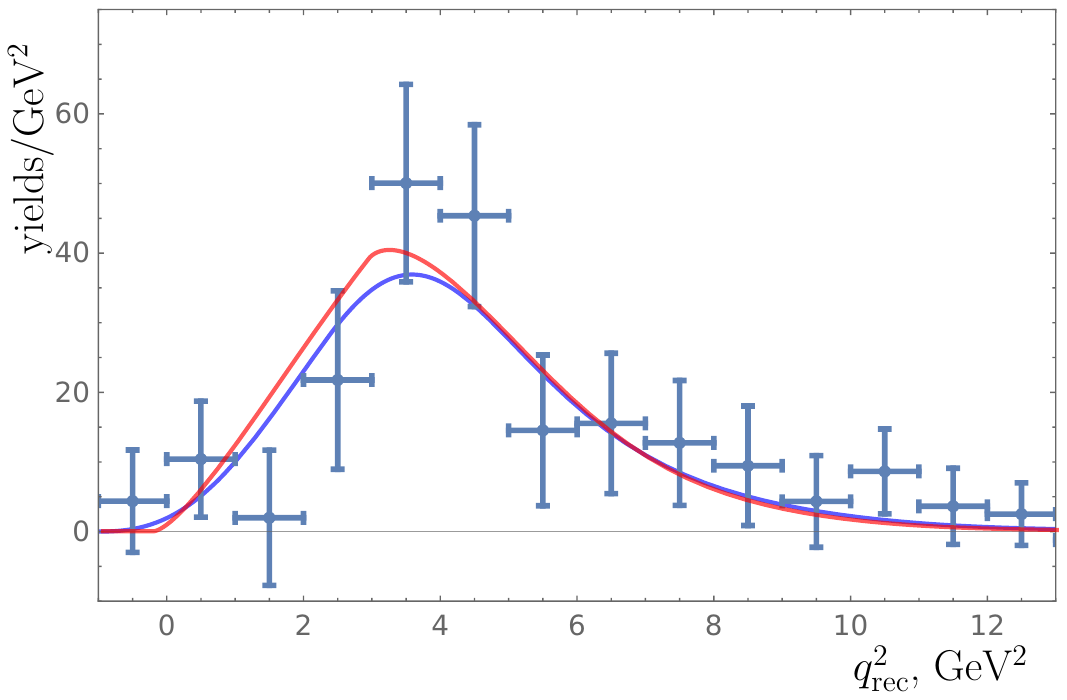}\\
 (c) & (d)
 \end{tabular}
 \caption{\label{Fig:2}
 Differential distributions in $B^+\to K^+ M_X$ decays:
 (a) Theoretical distribution for the S-scenario with DM parameters from (\ref{DMS}).
 (b) Theoretical distribution for the V-scenario with DM parameters from (\ref{DMV}).
 (c) Theoretical distribution for the SM.
 (d) Belle II excess events vs the predictions of the S-scenario (\textcolor{blue}{blue}) and the
 V-scenario (\textcolor{red}{red}), after applying the efficiency and smearing, for their
 DM parameters from (\ref{DMS}) and (\ref{DMV}).}
\end{figure}
\end{center}

Figure \ref{Fig:2} shows the differential distributions obtained in the SM and both S- and V-scenarios
with the DM parameters from (\ref{DMS}) and (\ref{DMV}). The results of fitting the Belle II excess
events in the differential distributions $d\Gamma(B\to KM_X)/dq^2$ within both S- and V-scenarios,
with their DM parameter values taken from (\ref{DMS}) and (\ref{DMV}), are also presented.

\subsection{Prediction of the event excess in $B\to K^* M_X$ and $B_s\to \phi M_X$ decays compared to the SM}
Within the mechanism residing at the core of top-philic DM scenarios, DM contributes in a similar way to
any~FCNC weak decay of a hadron containing the $b$-quark. Moreover, the DM parameters have been extracted
from fitting the Belle II data on $B\to K M_X$ and can be used for yielding parameter-free predictions
for other $B_{(s)}\to (P,V)M_X$~decays.

\begin{center}
 \begin{figure}[t]
 \begin{tabular}{cc}
 \includegraphics[width=8.5cm]{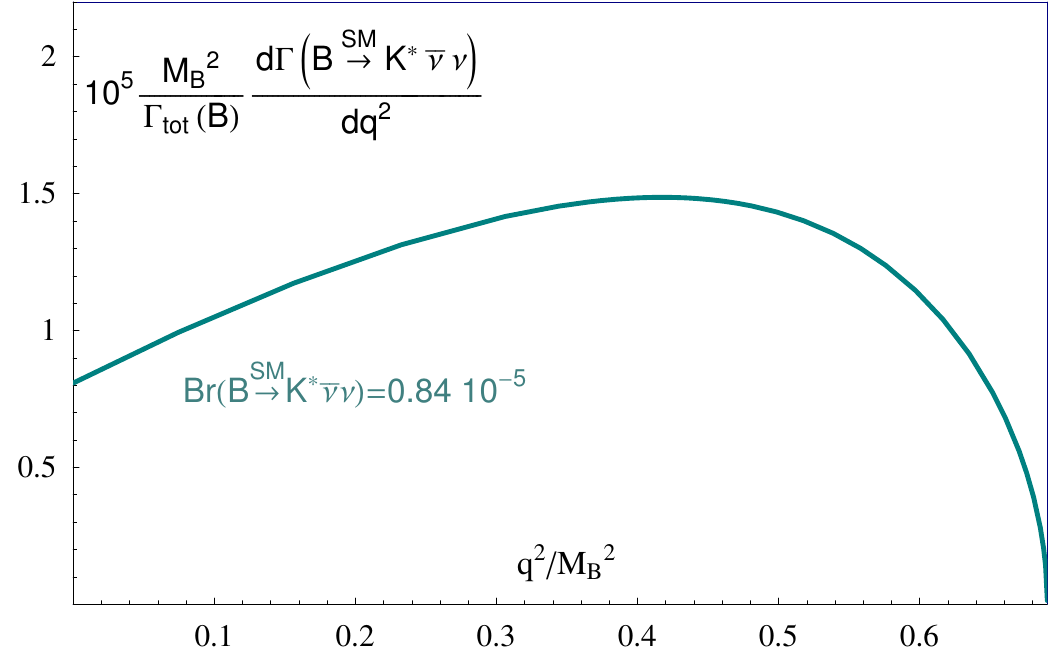}&
 \includegraphics[width=8.5cm]{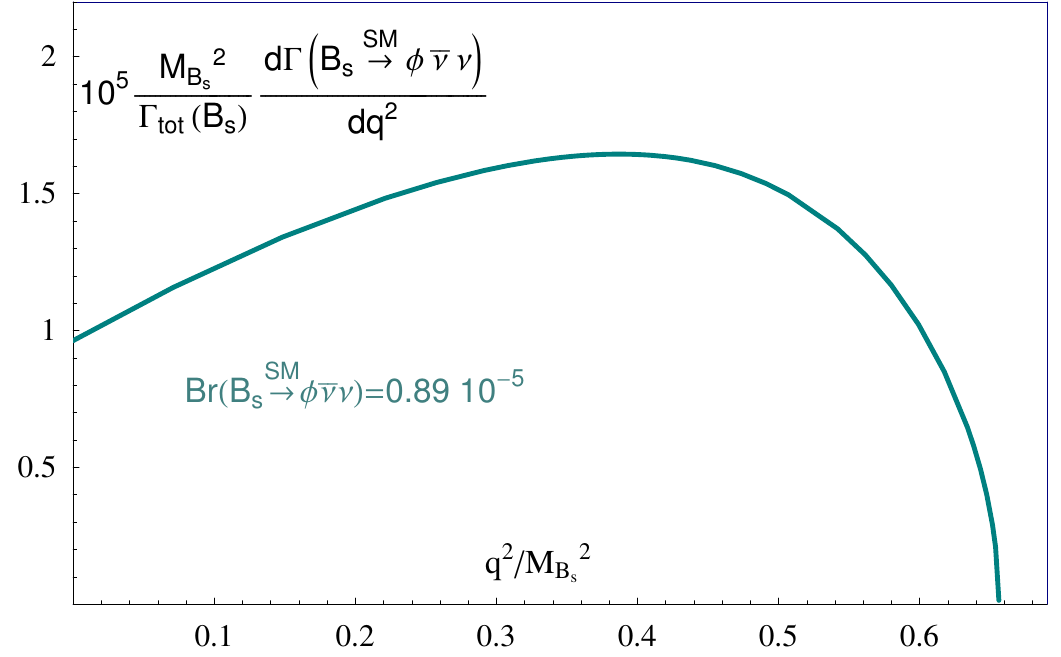}\\
 (a) & (b) \\
 \includegraphics[width=8.5cm]{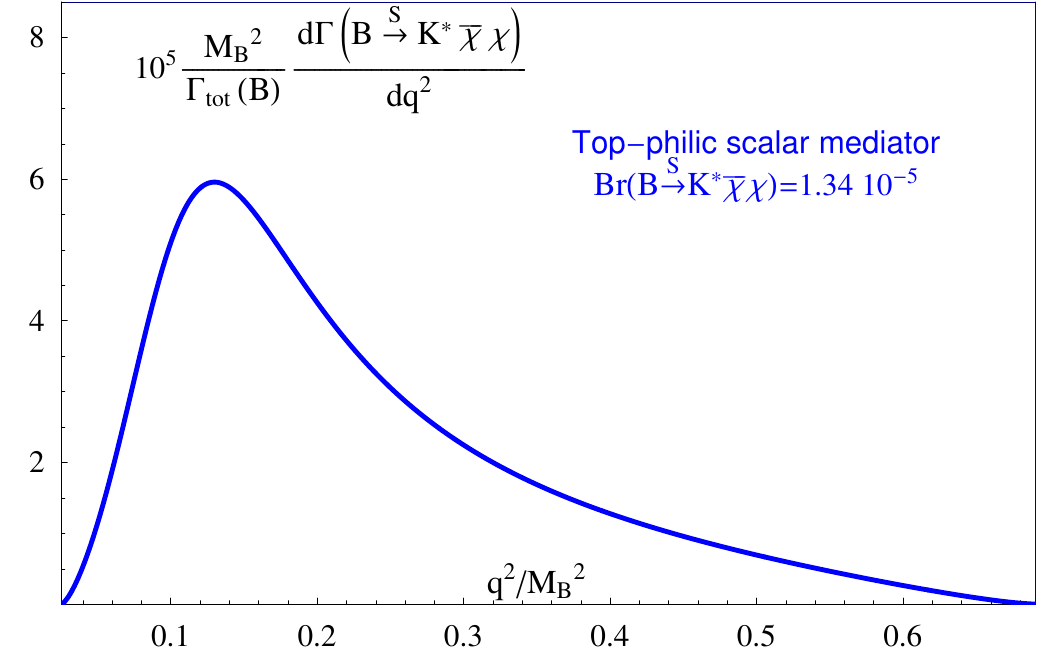}&
 \includegraphics[width=8.5cm]{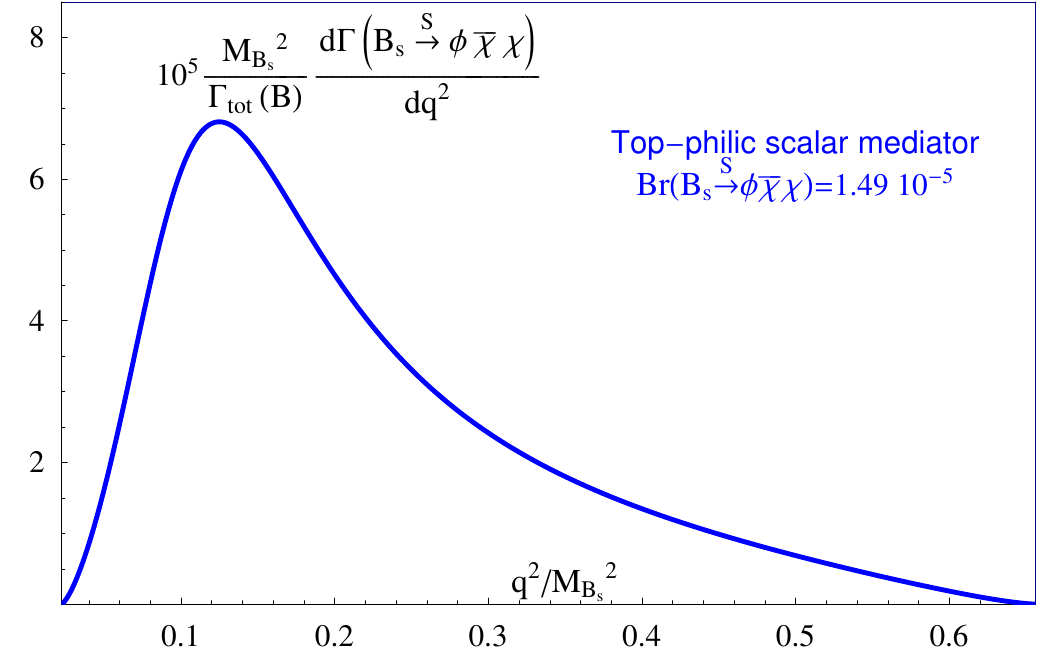}\\
 (c) & (d) \\
 \includegraphics[width=8.5cm]{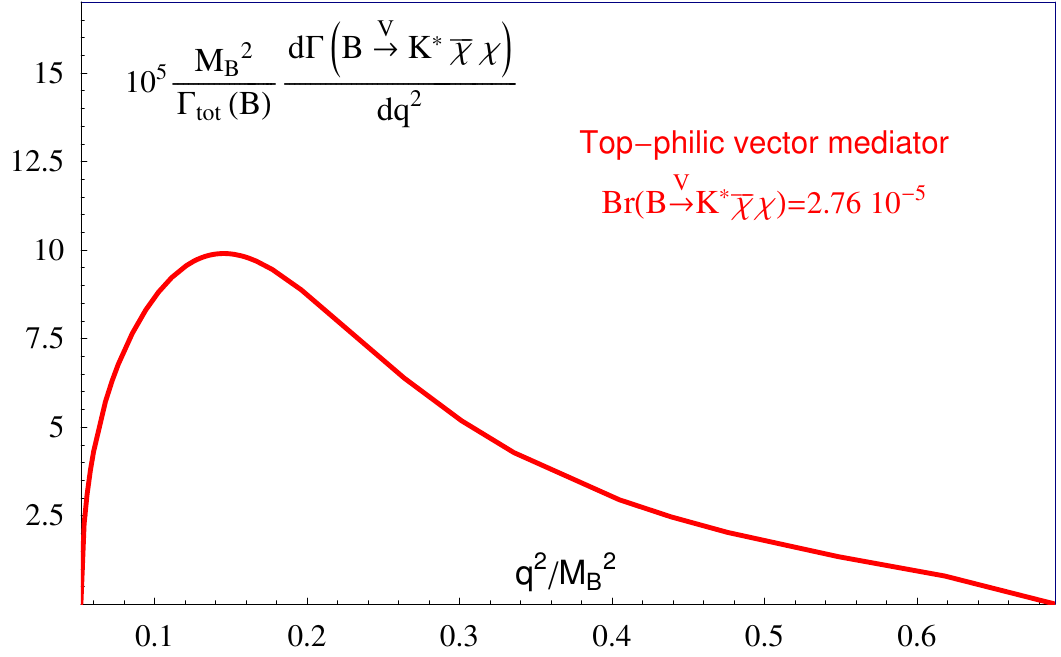}&
 \includegraphics[width=8.5cm]{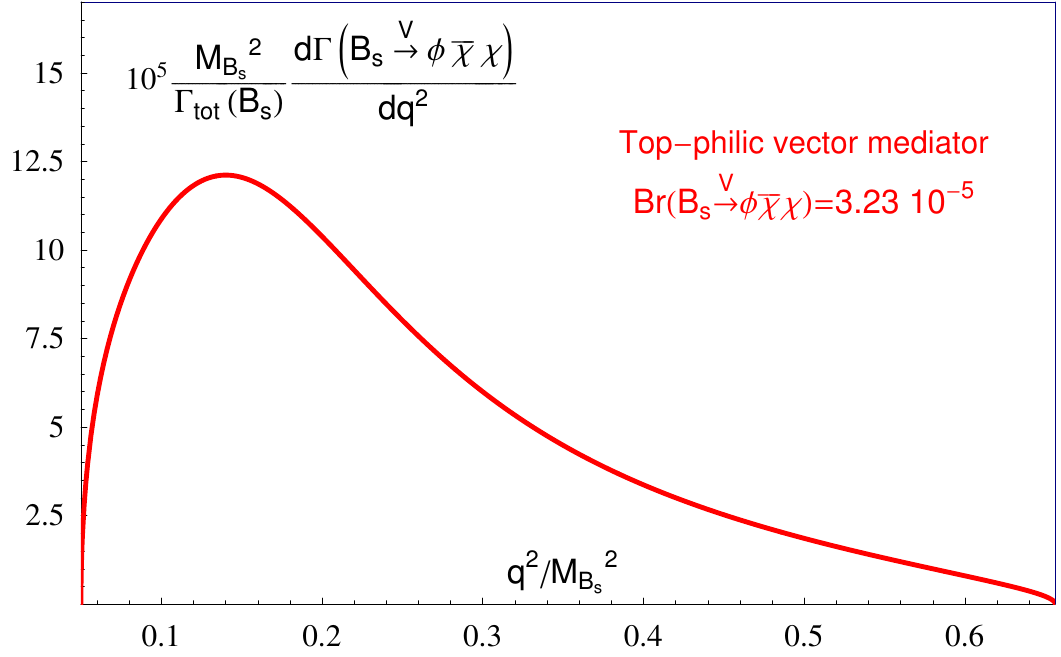}\\
 (e) & (f)
 \end{tabular}
 \caption{\label{Fig:3}
   Differential distributions in $B\to K^*$ decays (left column) and $B_s\to\phi$ decays (right column),
   in (a,b) the SM, (c,d) the DM S-scenario, and (e,f) the
   DM V-scenario.}
\end{figure}
\end{center}

Making use of the formulas from Appendix \ref{AppendixA}, the DM parameters given in Eqs.~(\ref{DMS})
and (\ref{DMV}), and the form-factor parameterizations in Appendix \ref{AppendixB}, we obtain the
differential and integrated branching fractions for the $B_s\to\phi M_X$~and $B\to K^* M_X$ decays
in Fig.~\ref{Fig:3}. It might be useful to indicate which form factors determine the
structure-dependent contributions to the decay modes of interest:
\begin{eqnarray}
 B\xrightarrow{SM}P\bar\nu\nu, \quad B\xrightarrow{V}P\bar\chi\chi:& \qquad &f_+^{B\to P}(q^2),\nonumber\\
 B\xrightarrow{SM}V\bar\nu\nu, \quad B\xrightarrow{V}V\bar\chi\chi:& \qquad &{\cal F}^{B\to V}(q^2),\nonumber\\
 B\xrightarrow{S}P\bar\nu\nu: &\qquad& f_0^{B\to P}(q^2),\nonumber\\
 B\xrightarrow{S}V\bar\nu\nu: &\qquad& A_0^{B\to V}(q^2),
\end{eqnarray}
where ${\cal F}$ represents the linear combination of the three form factors $A_1,A_2,V$
specified by Eqs.~(\ref{calF}) and (\ref{calF+}).

The integrated branching fractions can be found directly in the plots as well as in Eqs.~(\ref{BrB2Kstar})
and (\ref{BrBs2phi}). Evidently,~both the differential and the integrated branching fractions may serve
as clear discriminators between the SM, on the one hand, and the DM S- and V-scenarios, on the other hand.

To provide rigorous uncertainty estimates for these branching ratios turns out to be a very difficult task.
Taking into account that present realistic form-factor uncertainties are not likely to be
below 10\% (see the discussion in Appendix \ref{AppendixB}), conservative uncertainties in the branching
fractions are not likely to be below 20\% (although some publications provide more optimistic numbers).
A true reduction of the form-factor uncertainty depends on the availability of lattice QCD form factor
calculations in the full required range of $q^2$. On the positive side, however, note that the ratios
of the branching fractions may be predicted with better accuracy: For instance, the form-factor
dependent part completely drops out from the ratio of the differential distributions
$dBr(B\xrightarrow{V}K^*\bar\chi\chi)/dq^2$ and $dBr(B\xrightarrow{SM}K^*\bar\nu\nu)/dq^2$.

\subsection{Expected excess events in the Belle II $B\to K^* M_X$ data}
Let us demonstrate that the effects discussed above may be clearly visible in the experimental data.
As an example, we consider the decay $B\to K^* M_X$ under the conditions realized in the Belle II experiment.

Two features of the experimental analysis of Belle II should be recalled:
\begin{enumerate}
\item As described in detail in \citep{Belle-II:2023esi}, in order to increase the statistical
  sample the Belle II experiment analyzes partially reconstructed events. For events of this kind,
  the direction of the $B$ meson cannot be determined. Therefore, instead of $q^2\equiv(p_B-p_{K^*})^2$
  the variable $q_{\rm rec}^2$ is used:
 \begin{equation}
 q^2_{\rm rec}= E_B^2+M_{K^*}^2-2E_B E_{K^*},
 \end{equation}
 where $E_B$ and $E_{K^*}$ are the energies of the $B$ and $K^*$ mesons in the center-of-mass frame of
 the $B\bar B$-meson pair produced in $\Upsilon(4S)$ decays.
\item The Belle II experiment does not apply efficiency corrections to its data~\cite{Belle-II:2023esi}.
  Consequently, to obtain the expected number of events, we employ the $q^2$-dependent
  efficiency $\varepsilon(q^2)$ estimated in \cite{Fridell:2023ssf} (and provided by Fig.~2 of \cite{Fridell:2023ssf}).
\end{enumerate}
For the details of recalculating the differential distributions of interest in terms of $q^2_{\rm rec}$
(in other words, of effecting~the switch of variables $q^2\to q^2_{\rm rec}$), we refer to Ref.~\cite{blm2025}.

\begin{center}
\begin{figure}[ht]
\begin{tabular}{cc}
 \includegraphics[width=8.5cm]{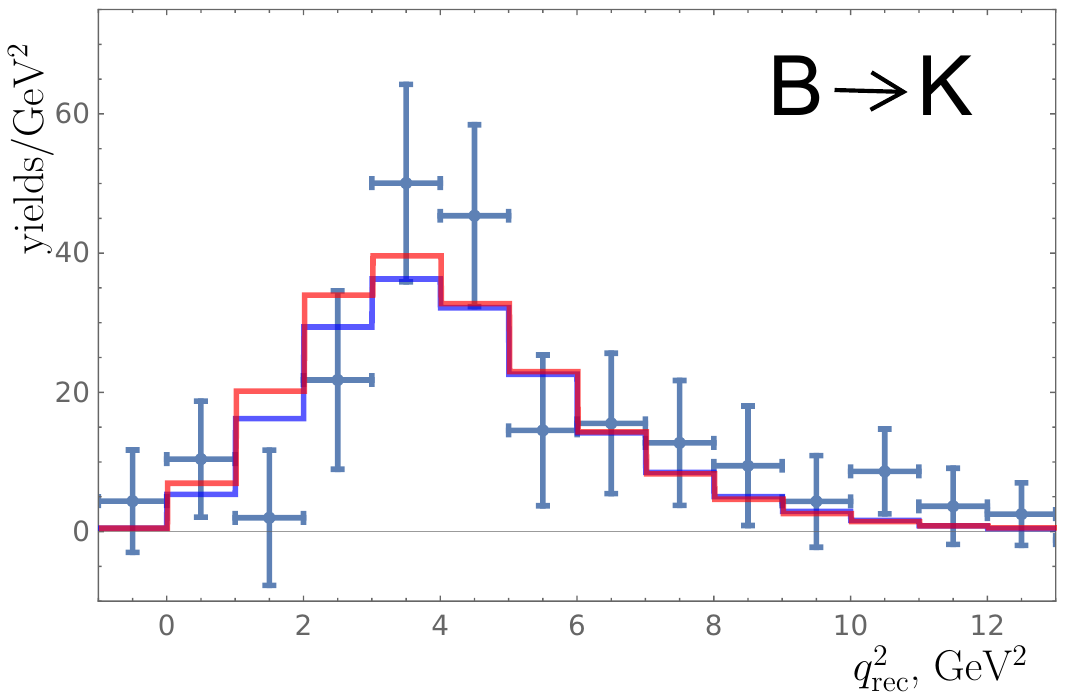} & \includegraphics[width=8.5cm]{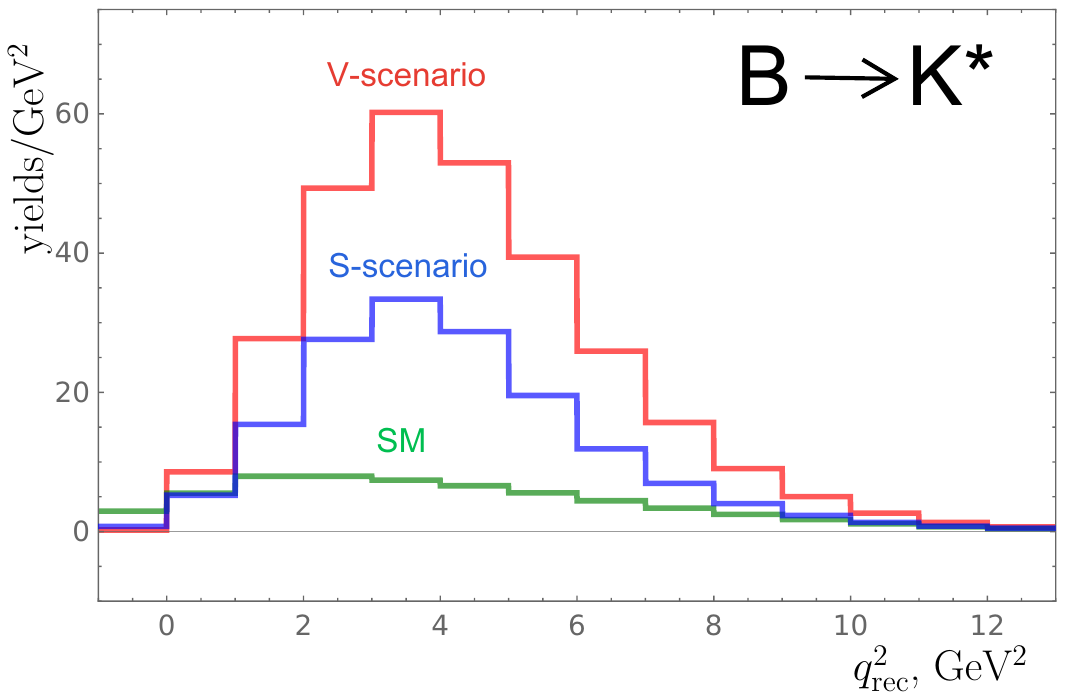}\\
 (a) & (b) 
\end{tabular}
 \caption{
   \label{Fig:4}
 The {\it observable} differential distributions vs $q_{\rm rec}^2$ binned as in Belle II data \cite{Belle-II:2023esi}: 
  SM (green), S-scenario (blue), and V-scenario (red).
  (a) $B\to K M_X$. The excess events measured by Belle II are shown.
  (b) $B\to K^* M_X$. For an illustration, the efficiency for $B\to K^* M_X$ is taken the same
  as the efficiency for $B\to K M_X$.}
\end{figure}
\end{center}

Figure \ref{Fig:4} provides our predictions for the differential distributions of the excess
events in the decay $B\to K^* M_X$, assuming the same detection efficiency for the $B\to K^*M_X$
and $B\to K M_X$ decays.

\section{\label{Sect:5} Summary and Conclusions}

We studied some consequences of the conjecture that the observed excess in the missing-energy
decay $B\to K M_X$ is due to the decay $B\xrightarrow{R}K\bar\chi\chi$ into two DM fermions
$\bar\chi\chi$ via a top-philic mediator field $R$. For the latter, two scenarios have been
considered: an S-scenario of a scalar mediator $R=R_s$ and a V-scenario of a vector mediator $R=R_v$.

The parameters of the DM sector in these two scenarios have been previously extracted from a
successful description of the Belle II data on $B\to K M_X$. Accordingly, we can provide
essentially parameter-free predictions for the DM contribution to the decays $B\to K^* M_X$ and $B_s\to \phi M_X$.

Our main findings are the following:

\begin{itemize}
\item In both S- and V-scenarios, the decay rates of $B\to K^* M_X$ and $B_s\to \phi M_X$ are
  strongly enhanced compared~to the SM result. Adopting for all parameters their central values,
  the differential distributions are shown in Fig.~\ref{Fig:3}.

For the corresponding integrated decay rates of $B\to K^* M_X$, we report
\begin{eqnarray}
 \label{BrB2Kstar}
 Br(B\xrightarrow{SM}K^*\bar\nu\nu) &=&0.84\times 10^{-5},\nonumber\\
 Br(B\xrightarrow{S}K^*\bar\chi\chi)&=&1.34\times 10^{-5},\nonumber\\
 Br(B\xrightarrow{V}K^*\bar\chi\chi)&=&2.76\times 10^{-5}.
\end{eqnarray}
For the integrated decay rate of $B_s\to\phi M_X$, we obtain
\begin{eqnarray}
 \label{BrBs2phi}
 Br(B_s\xrightarrow{SM}\phi\bar\nu\nu) &=&0.89\times 10^{-5},\nonumber\\
 Br(B_s\xrightarrow{S} \phi\bar\chi\chi)&=&1.49\times 10^{-5},\nonumber\\
 Br(B_s\xrightarrow{V} \phi\bar\chi\chi)&=&3.23\times 10^{-5}.
\end{eqnarray}
Obviously, the V-scenario entails a much stronger enhancement of the decay rate compared to
the SM expectation than the S-scenario.

For an illiustration, applying to $B\to K^*M_X$ events the same efficiency as
for $B\to K M_X$ events, Fig.~\ref{Fig:4} shows the expected number of excess
events in S- and V-scenarios in the sample where $B\to K M_X$ access has been
previoulsy reported. 

\item

  The uncertainty in these theoretical numbers is expected at the level of 20\%, mainly
  because of the uncertainty in the form factors $V, A_1,A_{12}$ governing the $B\to V$ transition.
  To acquire a rigorous control over these form-factor uncertainties is a notoriously difficult
  challenge. Looking at the broad picture of theoretical predictions~for these form factors
  (see Appendix \ref{AppendixB}), one can hardly imagine a realistic uncertainty in the form
  factors at the level of better than 10\%. This means that the uncertainty of the present
  theoretical predictions for the branching~ratios of $B_{(s)}\to V M_X$ decays ($V=K^*,\phi$)
  is expected to be at the level of 20\%. A sizeable reduction of the uncertainty may be
  expected with further progress in lattice QCD calculations.
\end{itemize}
Nevertheless, even with this sizeable uncertainty, {\it a clear discrimination between the DM S-
  and V-scenarios is possible as soon as the experimental data on $B\to K^* M_X$ and $B_s\to \phi M_X$ become available}.

\acknowledgments
The research of A.\,B.\ and D.\,M.\ was carried out within the scientific
program of the National Center for Physics and Mathematics,
Section 5  \emph{Particle Physics and Cosmology}. Stage 2026-2027.

\newpage
\appendix
\section{\label{AppendixA}$B\to (P,V)M_X$ decays in the SM and in dark-matter S- and V-scenarios}
To avoid confusion, we use the following notation throughout this Appendix:
$K$ denotes a generic pseudoscalar meson and $K^*$ denotes a generic vector meson.
To stay coherent with \cite{blm2025} and \cite{blm2026}, notations $\phi$ and $V$ are reserved for
pseudoscalar and vector DM mediator fields.

\subsection{$B\to (K,K^*)\bar\nu\nu$ decays in the SM}
For convenience, we reproduce here the formulas originally derived in \cite{colangelo1996,Melikhov:1997wp}.

\begin{itemize}
\item For a $B\to K\bar\nu\nu$ transition, the differential distribution reads ($M_1=M_B$ and $M_2=M_K$)
\begin{eqnarray}
\frac{d\Gamma(B\xrightarrow{SM}K\bar\nu\nu)}{dq^2}=\frac{c_L^2}{32\pi^3M_1^3}\lambda^{3/2}(M_1^2,M_2^2,q^2)|f_+|^2,
\end{eqnarray}
where
\begin{eqnarray}
c_L=\frac{G_F}{\sqrt{2}}\frac{\alpha_{\rm em}}{2\pi \sin^2(\theta_W)}V^*_{tb}V_{ts}X(x_t)
\end{eqnarray}
and $\lambda(a,b,c)=(a-b-c)^2-4bc$. Here, $f_+$ is the form factor describing the $B\to K$ weak decay amplitude.
\item For a $B\to K^*\bar\nu\nu$ transition, it has the form ($M_1=M_B$ and $M_2=M_{K^*}$)
\begin{eqnarray}
\label{dGammaSM}
\frac{d\Gamma(B\xrightarrow{SM}K^*\bar\nu\nu)}{dq^2}=\frac{c_L^2}{16\pi^3M_1^3}\lambda^{1/2}(M_1^2,M_2^2,q^2)\,{\cal F}(q^2),
\end{eqnarray}
where
\begin{eqnarray}
 \label{calF}
 {\cal F}(q^2)=32M_1^2M_2^2 A_{12}^2+q^2\left(\frac{\lambda(M_1^2,M_2^2,q^2)}{(M_1+M_2)^2}V^2+(M_1+M_2)^2A_1^2 \right)
\end{eqnarray}
and
\begin{eqnarray}
 \label{calF+}
 A_{12}=\frac{M_1+M_2}{16M_1M_2}\left((M_1^2-M_2^2-q^2)A_1-\frac{\lambda(M_1^2,M_2^2,q^2)}{(M_1+M_2)^2}A_2\right).
\end{eqnarray}
Here, $A_1,A_2,V$ are the standard form factors describing the amplitudes of $B\to K^*$ decays \cite{wsb}.
\end{itemize}
For the SM parameters, we use the following values (see Ref.~\cite{hpqcd2023}, Eq.~(15)): $X(x_t)=1.469(17)$, $\sin^2(\theta_W)=0.23121(4)$ and $1/\alpha_{\rm em}(M_Z)=127.952(9)$.

\subsection{$B\to (K,K^*)\bar\chi\chi$ decays in the S-scenario}
A popular model \cite{Batell:2009jf, Schmidt-Hoberg:2013hba}, referred to as $S$-scenario, involves an interaction of DM fermions $\chi$ with the top quark $t$ by exchange of a scalar-mediator field $\phi$, governed by the interaction Lagrangian
\begin{eqnarray}
 \mathcal{L_{\rm int}} = - \frac{y m_t}{v} \phi\, \bar t t - g_{\phi\chi\chi}\phi \bar \chi \chi,
 \label{eq:lagrangian}
\end{eqnarray}
where $v\simeq 246$ GeV is the Higgs vacuum expectation value and $y$ parametrizes the $\phi \bar tt$ coupling. The emerging effective Lagrangian encoding the FCNC vertex $b\to s\phi$ then reads
\cite{Batell:2009jf, Schmidt-Hoberg:2013hba}
\begin{eqnarray}
\label{Leff}
\mathcal{L}_{b \rightarrow s\phi} = g_{bs\phi}\,\phi\, \bar s_L b_R + {\rm h.c.}, \qquad
 g_{bs\phi}=\frac{y m_b}{v} \frac{3 \sqrt{2} G_F m_t^2 V^*_{ts} V_{tb}}{16 \pi^2}.
\end{eqnarray}
The $B\to (K,K^*)\bar\chi\chi$ transition proceeds via the mediator field $R$, see Fig.~\ref{Fig:1}.
One of the ingredients of the amplitude of interest, describing the $B\to K,K^*$ transition, has the form
\begin{eqnarray}
\label{FF}
 \langle K|\bar s_L b_R|B \rangle &=&\frac{1}{2}\langle K|\bar s (1-\gamma_5) b|B\rangle=
 \frac{1}{2}\langle K|\bar s b|B\rangle=
 \frac{1}{2} \frac{M_B^2-M_K^2}{m_b-m_s}f_0^{B\to K}(q^2),\nonumber\\
 \langle K^*|\bar s_L b_R|B \rangle &=& \frac{1}{2}\langle K^*|\bar s(1-\gamma_5) b|B \rangle=
 -\frac{1}{2}\langle K^*|\bar s \gamma_5 b|B \rangle
 =-i (\epsilon q)\frac{M_{K^*}}{m_b+m_s}A_0^{B\to K^*}(q^2),
\end{eqnarray}
with well-known dimensionless form factors $f_0^{B\to K}$ and $A_0^{B\to K^*}$ parameterizing the
amplitudes $\langle K|\bar s \gamma_\mu b|B\rangle$ and $\langle K^*|\bar s \gamma_\mu\gamma_5 b|B\rangle$ \cite{wsb}.
For quark masses in these formulas we take $m_b=4.2$ GeV and $m_s=0.1$ GeV. 

For the differential distributions in the $S$-scenario, we then obtained the following expressions \cite{blm2026}:
\begin{eqnarray}
\frac{d\Gamma(B\xrightarrow{S} K \bar\chi\chi)}{dq^2}&=&
\frac{\lambda^{1/2}(M_B^2,M_K^2,q^2)}{128 \pi^3 M_B^3}
\frac{(M_B^2-M_K^2)^2 |f_0^{B\to K}(q^2)|^2}{4(m_b-m_s)^2}
\frac{g_{bs\phi}^2\,g_{\phi\chi\chi}^2}{(M_\phi^2-q^2)^2+M_\phi^2\Gamma_\phi^2(q^2)}
 q^2 \left (1-\frac{4 m_\chi^2}{q^2}\right)^{3/2},\nonumber
\label{eq:dGdq2}\\
\frac{d\Gamma(B\xrightarrow{S} K^* \bar\chi\chi)}{dq^2}
&=&\frac{\lambda^{3/2}(M_B^2,M_{K^*}^2,q^2)}{128 \pi^3 M_B^3} \frac{|A_0^{B\to K^*}(q^2)|^2}{4(m_b+m_s)^2}
 \frac{g_{bs\phi}^2\,g_{\phi\chi\chi}^2}{(M_\phi^2-q^2)^2+M_\phi^2\Gamma_\phi^2(q^2)}
 q^2 \left(1-\frac{4 m_\chi^2}{q^2}\right)^{3/2},
\end{eqnarray}
where the $q^2$-dependent width of the mediator $\phi$ is calculated from the imaginary part of the fermion-loop diagram with scalar vertices in the form (cf. \cite{gs,nachtmann})
\begin{equation}
\label{Gamma}
\Gamma_\phi(q^2)= \left(\frac{q^2-4m_\chi^2}{M_\phi^2-4m_{\chi}^2}\right)^{\frac{3}{2}}\frac{M_\phi}{\sqrt{q^2}}\; \Theta(q^2-4m_\chi^2)\; \Gamma_\phi^0, \qquad \Gamma_\phi^0=\frac{g^2_{\phi\chi\chi}}{8\pi}M_\phi\left(1-\frac{4m_\chi^2}{M_\phi^2}\right)^{\frac{3}{2}}.
\end{equation}
We assume that $M_\phi> 2 m_\chi$ and, furthermore, that the mediator $\phi$ decays predominantly into the $\bar\chi\chi$ pair.

The form factors $f^{B\to K}_0$ and $A_0^{B\to K^*}$ are discussed in Appendix \ref{AppendixB}.

\subsection{$B\to (K,K^*)\bar\chi\chi$ decays in the V-scenario}
A widely discussed top-philic vector-mediator scenario is described by the Lagrangian
\cite{Langacker:2008yv,Cox:2015afa,Hu:2024xes}
\begin{eqnarray}
\mathcal{L}_{\rm int}=g_{Vtt}\,V^\mu\,\bar t \gamma_\mu (1+\gamma_5)t+g_{V\chi\chi}\, V^\mu\,\bar\chi \gamma_\mu \chi. \end{eqnarray}
Coupling of the mediator $V$ to the FCNC $b\to s$ current occurs through the penguin loop diagram of Fig.~1. Integrating out the heavy top and $W$ fields leads to the $b\to sV$ effective Lagrangian, which has the $V-A$ structure \cite{Inami:1980fz}
\begin{eqnarray}
\label{LbsV}
\mathcal{L}_{b \rightarrow sV}=g_{bs V} \bar s\gamma_\mu (1-\gamma_5)b\, V^\mu.
\end{eqnarray}
The Wilson coefficient $g_{bsV}\equiv \bar g_{bsV}(M_W)$ has the form
\begin{eqnarray}
 \label{gbsV}
 g_{bsV}&=&
 -\frac{g_{Vtt}}{16\pi^2}\frac{G_F}{\sqrt{2}}V_{tb}V_{ts}^*\,m_t^2\,F(x_t),\quad x_t=(m_t/M_W)^2,\nonumber \\
 &&F(x_t)=\frac{7-8x_t+x_t^2+2(4+(x_t-2)x_t)\log\,x_t}{(1-x_t)^2}=0.19.
\end{eqnarray}
The differential distributions $d\Gamma(B\xrightarrow{V} h\bar\chi\chi)/dq^2$, $h=K,K^*$,
were obtained in \cite{blm2026}:
\begin{eqnarray}
\label{dGammaDMV}
 \frac{d\Gamma(B\xrightarrow{V}h\bar\chi\chi)}{dq^2}=
 \frac{M_B^3}{192\pi^3}\sqrt{\lambda(1,r_h^2,\hat q^2)}\left(1+\frac{2m_\chi^2}{q^2}\right)\sqrt{1-\frac{4m_\chi^2}{q^2}}
 \frac{g_{bsV}^2 g_{V\chi\chi}^2}{(M_V^2-q^2)^2+\Gamma^2_V(q^2)M_V^2}\,\beta_h ,
\end{eqnarray}
with $\hat q^2=q^2/M_B^2$, $r_h=M_h/M_B$, and
\begin{eqnarray}
 \beta_K&=& |f_+^{B\to K}(q^2)|^2\lambda(1,\hat q^2,r_K^2), \\
 \label{betaKstar}
 \beta_{K^*}&=&\frac{2}{M_B^4}{\cal F}(q^2),
\end{eqnarray}
with ${\cal F}(q^2)$ given by Eq.~(\ref{calF}).
The form-factor dependent parts in Eq.~(\ref{dGammaSM}) corresponding to the SM and in Eq.~(\ref{betaKstar}) corresponding to the V-scenario are equal to each other.

Assuming that $M_V>2m_\chi$, the $q^2$-dependent width $\Gamma_V(q^2)$ is obtained as the imaginary part of the
two-point diagram with fermions in the loop and vector vertices \cite{nachtmann}, leading to
\begin{eqnarray}
 \label{GammaV}
 \Gamma_V(q^2)=\frac{M_V}{\sqrt{q^2}}
 \sqrt{\frac{q^2-4m_\chi^2}{M_V^2-4m_{\chi}^2}}\frac{q^2+2m_\chi^2}{M_V^2+2m_\chi^2}\Theta(q^2-4m_\chi^2)\; \Gamma_V^0,
\quad \Gamma_V^0=\frac{g^2_{V\chi\chi}M_V}{12\pi}\sqrt{1-\frac{4m_\chi^2}{M_V^2}}\left(1+\frac{2m_\chi^2}{M_V^2}\right).
\end{eqnarray}
These expressions involve the DM parameters $M_V$, $m_\chi$, $g_{bsV}$, and $g_{V\chi\chi}$, which may be extracted by fitting the data.

\newpage
\section{\label{AppendixB}The $B_{(s)}\to (K,K^*,\phi)$ form factors}

\begin{itemize}
\item For the $B\to K$ form factors $f_0$ and $f_+$, which are necessary to fit the Belle II results on $B\to KM_X$, we make use of the results from lattice QCD \cite{Bailey:2015dka} in the~form of rather convenient parameterizations proposed in \cite{ms2000}:
\begin{eqnarray}
\label{fplus}
f^{B\to K}_+(q^2) &=& \frac{0.335}{(1-q^2/M_{RV}^2)(1-0.58\,q^2/M_{RV}^2+0.03 (q^2/M_{RV}^2)^2)}, \\
\label{f0}
f^{B\to K}_0(q^2) &=& \frac{0.335}{1-0.648\,q^2/M_{RV}^2-0.17 (q^2/M_{RV}^2)^2},
\qquad M_{RV}=M_{B_s}(1^-)=5.415\mbox{ GeV}.
\end{eqnarray}
An uncertainty of 10\% may be safely assigned to the $B\to K$ form factors.
\item For the $B\to K^*$ and $B_s\to \phi$ form factors, we make use of the results from light-cone sum rules (LCSR) given in Table 14 of Ref.~\cite{Bharucha:2015bzk}, refitted to the form \cite{ms2000}
\begin{eqnarray}
\label{B2Vffs}
f(q^2)=\frac{f(0)}{(1-q^2/M_{\rm res}^2)\left(1-\sigma_1 q^2/M_{\rm res}^2+\sigma_2 (q^2/M_{\rm res}^2)^2\right)}.
\end{eqnarray}
The numerical results from \cite{Bharucha:2015bzk} are reproduced by our fit formulas with better than 0.5\% accuracy in the full decay region of $q^2$. The corresponding parameters are given in Tables \ref{Table:B2Kstar} and \ref{Table:Bs2phi}.
\end{itemize}

\begin{table}[h]
  \caption{\label{Table:B2Kstar}
    Fit results for the $B\to K^*$ form factors using the parametrization (\ref{B2Vffs}).}
  \begin{tabular}{|c|cccc|}
    \hline
  $B\to K^*$  & $\quad f(0)\quad $  & $\quad\sigma_1\quad$ &  $\quad\sigma_2\quad$ & $\quad M_{\rm res}[{\rm GeV}]\quad$ \\ \hline

$\quad A_0 \quad$    &  0.36  &  0.50     &  $-$0.03  &  5.366 \\ \hline
  $A_1$              &  0.27  &  $-$0.10  &  0.58     &  \\
  $A_{12}$           &  0.26  &  $-$0.41  &  1.10     &  5.829 \\
    $A_{2}$          &  0.23  &  0.45     &  0.1      &  \\ \hline
  $V $               &  0.34  &  0.54     &  $-$0.04  &  5.415\\ \hline
  \end{tabular}
\end{table}
\begin{table}[h]
  \caption{\label{Table:Bs2phi}
    Fit results for the $B_s\to \phi$ form factors using parametrization (\ref{B2Vffs}).}
  \begin{tabular}{|c|cccc|}
    \hline
  $B_s\to \phi$  & $\quad f(0)\quad $  & $\quad\sigma_1\quad$ &  $\quad\sigma_2\quad$ & $\quad M_{\rm res}[{\rm GeV}]\quad$ \\ \hline

$\quad A_0 \quad$    &  0.39  &  0.391    &  $-$0.2   &  5.366 \\ \hline
  $A_1$              &  0.30  &  $-$0.22  &  0.7      &  \\
  $A_{12}$           &  0.25  &  $-$0.46  & 1.36      &  5.829 \\
  $A_{2}$            &  0.256 &  0.45     &  0.1      &  \\ \hline
  $V $               &  0.39  &  0.53     &  $-$0.11  &  5.415\\   \hline
  \end{tabular}
\end{table}

To provide a realistic estimate of the uncertainties of the $B\to V$ form factors is not an easy task. In order to present an overview of the scattered results from different theoretical approaches, we collect a selected set of predictions for $V^{B_s\to \phi}(0)$ in Table~\ref{Table:ffs}. (For a more extensive comparison, we refer to Ref.~\cite{Ivanov:2011aa}, Table III). From the numbers given~in Table~\ref{Table:ffs}, it is hard to believe that the form factor $V$ is known with a few percent accuracy. So, we use the expressions for the form factors given by Tables \ref{Table:B2Kstar} and \ref{Table:Bs2phi} but assume the uncertainties at the level of 10\% for all the form factors.

\begin{table}[ht]
  \caption{\label{Table:ffs}
    Representative results for $B_s\to\phi$ form factors $V(0)$. The adopted abbreviations stand for the following approaches:
    RQM --- relativistic quark model; LCSR --- light-cone sum rules; Lat --- lattice QCD.}
  \begin{tabular}{|c|ccccccc|}
    \hline
    &  RQM \cite{ms2000}&  LCSR \cite{Ball:2004rg}&  RQM \cite{Ebert:2006nz} & RQM \cite{Ivanov:2011aa} &
    Lat \cite{Horgan:2013hoa}  &  LCSR \cite{Bharucha:2015bzk}  &  LCSR+Lat \cite{Bharucha:2015bzk}  \\
 \hline
    $V(0)$  &  0.44  & 0.454 $\pm$ 0.035  &  0.40  & 0.32  &  0.24 $\pm$ 0.07 &  0.39 $\pm$ 0.03&  0.36 $\pm$ 0.01\\
    \hline
  \end{tabular}
\end{table}


\end{document}